\documentclass[conference]{IEEEtran}
\IEEEoverridecommandlockouts

\usepackage{cite}
\usepackage{amsmath,amssymb,amsfonts}
\usepackage{algorithmic}
\usepackage{graphicx}
\usepackage{makecell}
\usepackage{multirow}
\usepackage{textcomp}
\usepackage{xcolor}
\usepackage[most,breakable]{tcolorbox}

\definecolor{boxcolor1}{RGB}{230,240,255} 
\definecolor{boxcolor2}{RGB}{255,240,230} 
\definecolor{boxcolor3}{RGB}{230,255,230} 
\definecolor{boxcolor4}{RGB}{255,230,255} 
\definecolor{boxcolor5}{RGB}{255,255,230} 
\definecolor{boxcolor6}{RGB}{230,255,255} 
\definecolor{grey}{RGB}{240,240,240} 

\definecolor{darkgreen}{rgb}{0, 0.5, 0}

\newtcolorbox{questionbox}[1][]{
  colback=#1,
  colframe=black,
  boxrule=1pt,
  arc=5pt,
  outer arc=5pt,
  enhanced,
  breakable
}

\def\BibTeX{{\rm B\kern-.05em{\sc i\kern-.025em b}\kern-.08em
    T\kern-.1667em\lower.7ex\hbox{E}\kern-.125emX}}

\begin{document}

\title{AstroMLab 2: AstroLLaMA-2-70B Model and Benchmarking Specialised LLMs for Astronomy}

\author{
    \IEEEauthorblockN{Rui Pan$^*$\thanks{* Equal contribution.}}
    \IEEEauthorblockA{
        \textit{Department of Computer Science} \\
        \textit{University of Illinois Urbana-Champaign}\\
        Urbana, USA \\
        ruip4@illinois.edu
    }
    \and
    \IEEEauthorblockN{Tuan Dung Nguyen$^*$}
    \IEEEauthorblockA{
        \textit{Department of Computer and Information Science} \\
        \textit{University of Pennsylvania}\\
        Philadelphia, PA 19104, USA \\
        joshtn@seas.upenn.edu
    }
    \and
    \IEEEauthorblockN{Hardik Arora}
    \IEEEauthorblockA{
        \textit{Indian Institute of Technology} \\
        Patna, India \\
        hardikarora1707@gmail.com
    }
    \and
    \IEEEauthorblockN{Alberto Accomazzi}
    \IEEEauthorblockA{
        \textit{Center for Astrophysics} \\
        \textit{Harvard University}\\
        Cambridge, USA \\
        aaccomazzi@cfa.harvard.edu
    }
    \and
    \IEEEauthorblockN{Tirthankar Ghosal}
    \IEEEauthorblockA{
        \textit{National Center for Computational Sciences} \\
        \textit{Oak Ridge National Laboratory}\\
        Oak Ridge, TN, USA \\
        ghosalt@ornl.gov
    }
    \and
    \IEEEauthorblockN{Yuan-Sen Ting}
    \IEEEauthorblockA{
        \textit{Department of Astronomy} \\
        \textit{The Ohio State University}\\
        Columbus, USA \\
        ting.74@osu.edu
    }
}

\maketitle

\begin{abstract}
Continual pretraining of large language models on domain-specific data has been proposed to enhance performance on downstream tasks. In astronomy, the previous absence of astronomy-focused benchmarks has hindered objective evaluation of these specialized LLM models. Leveraging a recent initiative to curate high-quality astronomical MCQs, this study aims to quantitatively assess specialized LLMs in astronomy. We find that the previously released AstroLLaMA series, based on LLaMA-2-7B, underperforms compared to the base model. We demonstrate that this performance degradation can be partially mitigated by utilizing high-quality data for continual pretraining, such as summarized text from arXiv. Despite the observed catastrophic forgetting in smaller models, our results indicate that continual pretraining on the 70B model can yield improvements. However, the current supervised fine-tuning dataset still constrains the performance of instruct models. In conjunction with this study, we introduce a new set of models, AstroLLaMA-3-8B and AstroLLaMA-2-70B, building upon the previous AstroLLaMA series.
\end{abstract}


\section{Introduction}

Large Language Models (LLMs), such as GPT~\cite{radford2019} and LLaMA~\cite{touvron2023a,touvron2023b,dubey2024}, have demonstrated remarkable capabilities across a wide range of tasks, both general and domain-specific \cite{nelson2019, brown2020, kaplan2020, achiam2023}. This versatility has paved the way for extending their utility beyond natural language processing and capable of performing domain-specific research tasks \cite{azerbayev2023,yoshikawa2023,jablonka2024,lei2024,wang2024b,wang2024a,ziems2024}, planetary science~\cite{yang2024} and astronomy~\cite{sun2024a}. These applications encompass not only conventional tasks such as recommender systems \cite{geng2022,chu2023,zhao2023,vats2024}, and knowledge graphs \cite{kau2024,sun2024a} but also more advanced logical tasks, including conducting end-to-end research through LLM agents to expedite scientific discovery \cite{boiko2023,bran2023,ramos2024,sun2024b}.

However, even highly capable general models such as GPT and LLaMA face limitations in executing intricate tasks. The API charges for closed-source models such as GPT can be prohibitive for the large-scale deployment of multi-LLM-based agents. Additionally, open-weight models tend to focus on general, everyday concepts rather than the specialized, infrequent content often found in academic research, potentially limiting their effectiveness in developing research-based LLM agents. Furthermore, infrequent updates to LLMs' training datasets may result in delays in incorporating the most recent findings, which is often compounded by the lack of a unified data repository in many fields.

These limitations have led to proposals for training specialized LLMs for individual domains, with the goal of creating reliable and specialized open-source LLMs that can be adapted as part of the aforementioned research agents. In the field of astronomical research, several models have been further pre-trained based on LLaMA-2-7b~\cite{nguyen2023,perkowski2024} and Mistral~\cite{dehaan2024}. However, efforts to develop LLMs for astronomy have been hampered by the lack of comprehensive benchmarking to quantify the capabilities of these models.

While various benchmarking efforts have been conducted in other areas \cite{wang2019,hendrycks2020,lin2021,liang2022,srivastava2023}, these general benchmarks fall short in evaluating the specific skills required for astronomical research. Ting et al. 2024 \cite{ting2024} has demonstrated that even flagship models deemed comparable in other benchmarks can differ in specific astronomy benchmarks, with variations of up to two to three orders of magnitude in terms of cost-efficiency. The benchmarking dataset developed in \cite{ting2024} provides the first objective metric for evaluating some of the released specialized LLMs quantitatively.

In this study, we extend beyond benchmarking existing models, notably AstroLLaMA \cite{nguyen2023,perkowski2024}, which is based on LLaMA-2-7B. We release a new set of models\footnote{https://huggingface.co/AstroMLab} building upon the previous AstroLLaMA series: AstroLLaMA-3-8B and AstroLLaMA-2-70B. This work aims to establish a robust benchmark that accurately assesses the capabilities of LLMs in astronomical research, particularly their ability to recall astronomical facts and make broad inferences based on current astronomical consensus. 

\section{Existing Specialized LLMs for Astronomy}

At the commencement of this study (April-July 2024), two primary specialized LLMs for astronomy were available: AstroLLaMA~\cite{nguyen2023} and its successor, the AstroLLaMA-chat model~\cite{perkowski2024}\footnote{AstroMLab comprises some core members of the original AstroLLaMA team and the authors of this study.}. It's worth noting that during the course of our research, we became aware of another specialized LLM, CosmoSage, introduced by \cite{dehaan2024}. However, our analysis focuses primarily on the models available at the study's inception.

Both AstroLLaMA models are based on Meta's LLaMA-2-7b architecture, fine-tuned using astronomical literature from arXiv's astro-ph category. The initial AstroLLaMA~\cite{nguyen2023} employed Parameter-Efficient Fine-Tuning (PEFT) with LowRank Adaptation (LoRA), focusing on abstracts from 326,238 astronomy papers (April 1992 to July 2023). The subsequent AstroLLaMA-Chat~\cite{perkowski2024} expanded this approach by including introductions and conclusions of papers, and incorporating domain-specific dialogue datasets such as LIMA~\cite{zhou2024}, Open Orca~\cite{openorca2023,mukherjee2023,longpre2023}, and UltraChat~\cite{ding2023}.

\section{Extending AstroLLaMA: AstroLLaMA-2-70B and AstroLLaMA-3-8B}

Building upon the foundation laid by previous AstroLLaMA models and incorporating some of the latest models, including LLaMA-3, we extended our model series to include LLaMA-2-70B and AstroLLaMA-3-8B. This expansion provides us with a broader baseline to determine optimal strategies for continual pretraining. We chose these models because the LLaMA-2-70B model has shown remarkable success in various benchmark datasets, while AstroLLaMA-3-8B has outperformed LLaMA-2-70B in our astronomy benchmarking \cite{ting2024}.

The training process for the new AstroLLaMA series follows two steps: continual pretraining (CPT) on the dataset, followed by specialized fine-tuning (SFT). For our CPT dataset, we applied the same dataset as in~\cite{perkowski2024} for direct comparison. This dataset comprises all arXiv papers from the astro-ph category, from the inception of arXiv up to July 2023. We extracted the abstract, introduction, and conclusion sections, denoting models trained on this dataset as ``AIC" (e.g., AstroLLaMA-3-8B-AIC, AstroLLaMA-2-70B-AIC), as these sections often contain the most pertinent information for understanding astronomy concepts. To normalize the notation for readability, we will rename the previous AstroLLaMA models as AstroLLaMA-2-7B-Abstract for the one from \cite{nguyen2023} and AstroLLaMA-2-7B-AIC for the one from \cite{perkowski2024}, based on the training dataset.

We also note that the original AIC dataset, derived purely from arXiv LaTeX sources, underwent extensive algorithmic cleaning in \cite{perkowski2024}. However, we found that some methods did not fully provide excellent data quality. As part of a broader effort to be published in the future from AstroMLab, we have turned to performing optical character recognition (OCR)  on all PDF files from arXiv (downloaded through ADS), extending from arXiv's inception up to January 2024. We used the Nougat tool~\cite{blecher2023} for the OCR, transcribing these papers into text. Further, for this study, we also used Qwen-2-8B and LLaMA-3.1-8B to summarize the PDF files, reducing them to about 1,000-4,000 tokens, roughly equivalent to the AIC set in training tokens. This approach aims to incorporate detailed knowledge beyond the AIC while maintaining affordable CPT for this study. We label this model as AstroLLaMA-3-8B-Summary.

To efficiently manage the CPT of these large models, we employed the LMFlow framework~\cite{diao2024}, which incorporates several advanced techniques to accelerate training. These optimizations were crucial, particularly for the LLaMA-2-70B model. For AstroLLaMA-3-8B, we used the following hyperparameters: learning rate of $2\times 10^{-5}$, total batch size of 96, maximum token length of 512, warmup ratio of 0.03, and no gradient accumulation, and the use of the bf16 format. For AstroLLaMA-2-70B, we adjusted some parameters: learning rate of $2\times 10^{-5}$, total batch size of 160, maximum token length of 2048, warmup ratio of 0.03. Both models used a cosine decay schedule~\cite{loshchilov2016} for learning rate reduction.

To transform the base models into helpful assistant models, we performed SFT to create ``chat/instruct" versions. We used the same conversation dataset as AstroLLaMA-Chat~\cite{perkowski2024}, which includes 10,356 astronomy-centered conversations generated from arXiv abstracts by GPT-4, the full content of LIMA~\cite{zhou2024}, 10,000 samples from Open Orca~\cite{openorca2023,mukherjee2023,longpre2023}, and 10,000 samples from UltraChat~\cite{ding2023}. We note that this SFT set is not highly tuned to astronomy Q\&A, with only one-third of the samples being astronomy-focused. For the SFT process, the learning rate was set to $3 \times 10^{-7}$, with one training epoch, a total batch size of 48, a maximum token length of 2048, a warmup ratio of 0.03, and we also used a cosine decay schedule. 

The computational demands for this project were substantial. CPT cost about 32 A100 GPU A100 hours for the 8B models and about 2,000 A100 GPU hours for the 70B models; in all cases, we only trained for one epoch. SPT required about 12 A100 GPU hours for the 8B model and 100 GPU hours for the 70B model. Inference time for the 70B models, particularly for fully instruct Q\&A (output up to 512 tokens), cost about 64 A100 GPU hours for all 4,425 multiple-choice questions (MCQs).

\section{Benchmark MCQ Datasets}

For our evaluation, we adopt the comprehensive benchmarking dataset specifically designed to assess LLMs in the context of astronomical research, as described in \cite{ting2024}. This dataset is derived from the Annual Review of Astronomy and Astrophysics, widely regarded as the most authoritative journal in the field, publishing extensive reviews by world-leading experts. Each article in this journal provides a comprehensive summary of state-of-the-art research in a specific astronomical subfield. This approach ensures a broad, non-myopic view of each topic, with contributions from authors recognized as world leaders in their respective areas.

To generate the MCQ dataset, we leveraged the long-context capabilities of Gemini-1.5-Pro-001~\cite{team2023gemini}. Through carefully curated few-shot examples, precise prompting, and iterative refinement with human input, we extracted high-quality MCQs from these comprehensive reviews. The MCQ extraction process adhered to several key principles. Questions were designed to be independent of specific article results, ensuring their standalone usability. The focus was on broad consensus and current state-of-the-art knowledge, often integrating insights from various subfields – a challenge that general LLMs might struggle with. Additionally, answer options were crafted to be of equal length, preventing easy elimination based on superficial characteristics.

Our process involved 885 ARAA articles, generating five questions per article, each with four answer options. This approach yielded a total of 4,425 questions, encompassing a wide spectrum of astronomical topics and concepts. The difficulty level of these questions is comparable to that of general exams in Ph.D. programs in astrophysics within the United States educational system. Some examples can be found in Appendix~\ref{AppendixA}.

\begin{table*}[htbp]
  \centering
  \begin{minipage}{\textwidth}
  \centering
  \caption{Performance of LLaMA and AstroLLaMA models on astronomy MCQ benchmarks. Scores show the fraction of accurate answers provided, for three benchmarking methods as described in the text. The arrows indicate if the AstroLLaMA models are performing better (\textcolor{darkgreen}{$\Uparrow$}), worse (\textcolor{red}{$\Downarrow$}), or similarly (\textcolor{orange}{$\Rightarrow$}) compared to the native LLaMA series models.}
  \label{tab:main}
  \begin{tabular}{lccccc}
  \hline\\[-0.1cm]
  \textbf{Model} & \textbf{Full Instruct (\%)} & \textbf{Token Prediction} & \textbf{Token Prediction} & \textbf{Source} & \textbf{Reference} \\
   &  & \textbf{(Instruct Model) (\%)} & \textbf{(Base Model) (\%)} & & \\[0.2cm]
  \hline\\[-0.1cm]
  \multicolumn{6}{l}{\textbf{LLaMA-2 Series (7B Parameters)}} \\[0.05cm]
  LLaMA-2-7B & 50.3 & 62.6 & 51.3 & Meta & \cite{touvron2023b} \\[0.2cm]
  
  \multicolumn{6}{l}{\textbf{AstroLLaMA-2 Series (7B Parameters)}} \\[0.05cm]
  AstroLLaMA-2-7B-AIC$^b$ & 41.4 \textcolor{red}{$\Downarrow$} & 47.2 \textcolor{red}{$\Downarrow$} & 44.3 \textcolor{red}{$\Downarrow$} & uTBD & \cite{perkowski2024} \\
  AstroLLaMA-2-7B-Abstract$^a$ & - & - & 43.5 \textcolor{red}{$\Downarrow$} & uTBD & \cite{nguyen2023} \\[0.2cm]
  \hline\\[-0.1cm]
  
  \multicolumn{6}{l}{\textbf{LLaMA-3 Series (8B Parameters)}} \\[0.05cm]
  LLaMA-3-8B & 72.9 & 73.6 & 72.0 & Meta & \cite{dubey2024} \\[0.2cm]
  
  \multicolumn{6}{l}{\textbf{AstroLLaMA-3 Series (8B Parameters)}} \\[0.05cm]
  AstroLLaMA-3-8B-AIC$^b$ & 61.8 \textcolor{red}{$\Downarrow$} & 68.4 \textcolor{red}{$\Downarrow$} & 71.9 \textcolor{orange}{$\Rightarrow$} & AstroMLab & This Study \\
  AstroLLaMA-3-8B-Summary$^c$ & 69.0 \textcolor{red}{$\Downarrow$} & 70.9 \textcolor{orange}{$\Rightarrow$} & 72.3 \textcolor{orange}{$\Rightarrow$} & AstroMLab & This Study \\[0.2cm]
  \hline\\[-0.1cm]
  
  \multicolumn{6}{l}{\textbf{LLaMA-2 Series (70B Parameters)}} \\[0.05cm]
  LLaMA-2-70B & 70.7 & 71.4 & 73.9 & Meta & \cite{touvron2023b} \\[0.2cm]
  
  \multicolumn{6}{l}{\textbf{AstroLLaMA-2 Series (70B Parameters)}} \\[0.05cm]
  AstroLLaMA-2-70B-AIC$^b$ & 64.7 \textcolor{red}{$\Downarrow$} & 75.4 \textcolor{darkgreen}{$\Uparrow$} & 76.0 \textcolor{darkgreen}{$\Uparrow$} & AstroMLab & This Study \\[0.2cm]
  \hline\\[-0.1cm]
  \end{tabular}
  
  \end{minipage}
  
  \vspace{0.5em}
  \begin{flushleft}
  \footnotesize
  $^a$ Abstract: performed CPT on the abstract of the astro-ph arXiv data.\\
  $^b$ AIC: performed CPT on the abstract, introduction and conclusion of the astro-ph arXiv data.\\
  $^c$ Summary: performed CPT on the full text summary from astronomy astro-ph papers generated with LLaMA-3.1-8B and Qwen-2-8B.
  \end{flushleft}
\end{table*}

\section{Inference Methodology}

To quantify the performance of specialized LLMs for astronomy, we employed three distinct benchmarking methods.

\subsection{Full Instruct Benchmarking Method}

The primary method is the ``full instruct" benchmarking approach, as detailed in \cite{ting2024}. This method provides a robust framework for evaluating the models' capabilities to perform as assistant models, where question answering is conducted through natural conversation. It aims to assess the models' ability to engage in multi-turn reasoning and follow nuanced instructions, mirroring the requirements of practical research applications in astronomy.

In implementing this method, we utilized the Instruct versions of the models, with prompt details available in Appendix~\ref{appendixB}. We incorporated chain-of-thought prompting, requiring models to explain their reasoning alongside their answers. This approach has been demonstrated to improve accuracy in MCQ tasks across various models \cite{wei2022,zhou2022}. For proprietary models, we adhered to the default instructions (including temperature settings) provided in their API documentation, while for open-weights models, we followed the guidelines from the correspoding Hugging Face release.

The full instruct method presented some challenges, particularly with weaker or earlier open-weights models that showed inconsistent adherence to the specified output format. To address this, we implemented a preliminary regex to extract answers in most cases. In the rare instances where this failed, we employed a GPT-4o model to interpret the intended answer from the model's explanation. This approach ensured a consistent evaluation across all models, regardless of their output format consistency.

\subsection{Base Model Token Benchmarking Method}

Upon observing that the full instruct models of AstroLLaMA consistently underperformed, we sought to understand whether this degradation occurred during the CPT process (i.e., when improving next-token prediction) or during the SFT process (i.e., when enhancing instruction-following capabilities). To dissect this difference, we implemented a second benchmarking approach: the base model (next-)token method. This method is designed to evaluate pre-trained AstroLLaMA base models before SFT, thus eliminating any influence from the SFT. Since base models lack instruction-following capabilities, the most appropriate evaluation approach is to test their ability to complete text.

In this method, we present the model with the question and all four options, followed by the prompt ``Answer:". We then analyze the logits of the next token to determine the most probable answer (A, B, C, or D). To provide context for the task and improve performance, we employ a two-shot approach by including two example questions with their correct answers in the prompt before presenting the actual test question. This approach gives the model a clear pattern to follow, which is particularly beneficial for base models not specifically fine-tuned for instruction following.

Our evaluation accounts for variations in token representation across different models. For instance, some models may represent answer choices as ``A", ``B", ``C", ``D", while others use `` A", `` B", `` C", `` D" (note the leading space). Our code dynamically identifies the correct token representation for each model by examining the top ten tokens in the model's output. This flexibility ensures accurate evaluation across various model architectures and tokenization schemes. To ensure reproducibility and consistency in our results, we set the temperature to 0.0 during inference for this benchmarking method.

\subsection{Instruct Model Token Benchmarking Method}

To complete our evaluation approach, we introduced a third method: the instruct model (next-)token benchmarking method. This approach applies the token prediction technique as described in the previous section to the instruct models that have undergone SFT. This benchmarking method allows us to investigate whether SFT, in cases of apparent performance degradation, affects the model's underlying knowledge structure or primarily impacts its instruction-following capability.

\section{Results}

\begin{figure*}[htbp]
\centering
\includegraphics[width=1.0\textwidth]{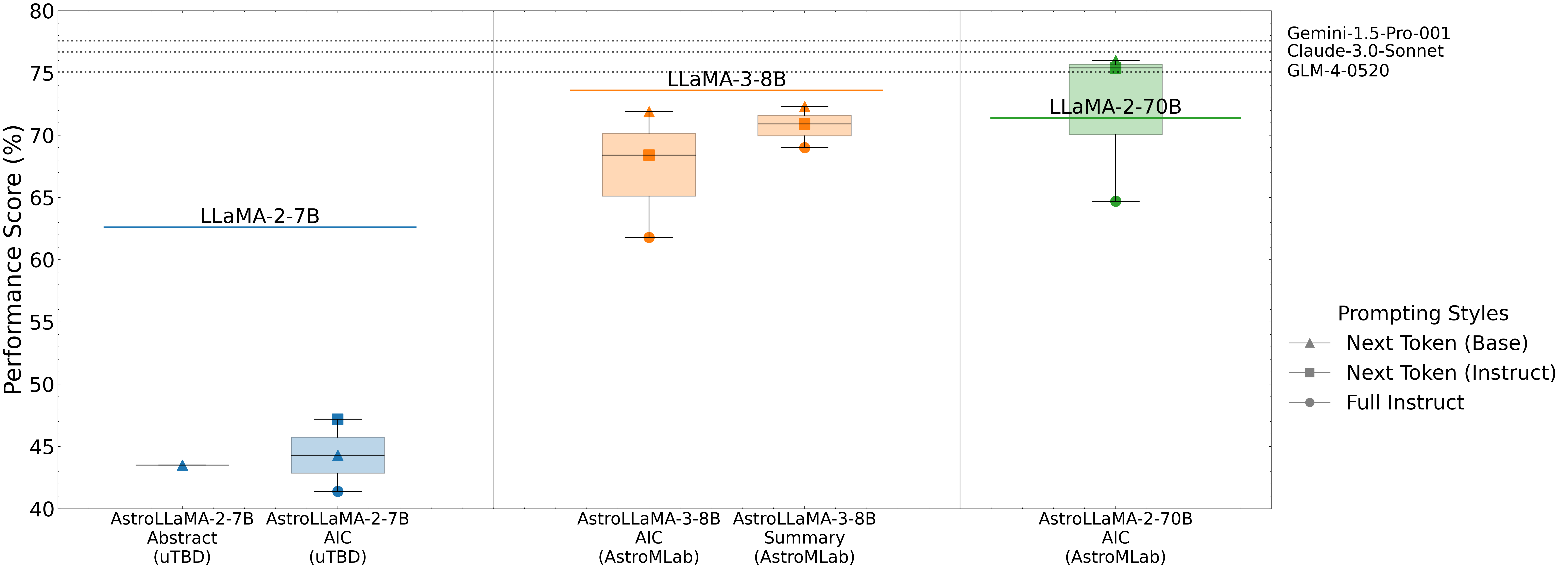}
\caption{Performance comparison of baseline LLaMA models and their specialized AstroLLaMA counterparts on astronomy MCQ benchmarks. Scores are shown as percentages for different prompting styles: full instruction-following, next-token prediction (instruct model), and next-token prediction (base model), represented by three different symbols. Horizontal lines indicate the full instruct scores of native LLaMA models for each corresponding series the AstroLLaMA is trained on. The existing AstroLLaMA-2-7B shows a notable decrement in ability. The AstroLLaMA-3-8B in this study mitigates that problem; however, we find that training on astro-ph data alone fails to improve the performance of the 8B models. This study introduces the first specialized LLM in astronomy at the 70B parameter level. The AstroLLaMA-2-70B outperforms the native LLaMA-2-70B models, demonstrating that training on astro-ph data can improve knowledge recall performance at the 70B level. Notably, across all models, the instruct versions, especially when evaluated using the full instruct benchmarking method, perform worse than the next-token prediction task. This suggests that the current lack of astronomy-focused Q\&A for SFT poses challenges in developing a useful assistant model from the specialized base model.}
\label{fig:main}
\end{figure*}

We evaluate the performance of various models using the fraction of correctly answered MCQs as our benchmark scores. The results are presented in Table~\ref{tab:main} and Figure~\ref{fig:main}. For the baseline methods, the scores from the full instruct benchmarking method align with those reported in \cite{ting2024}, as we employ the same dataset and inference protocol.

Our analysis reveals that the scores for the native LLaMA models are remarkably consistent across all benchmarking methods - whether using next-token prediction with the base model or the instruct model, or the full instruct assistant mode. The sole exception is the LLaMA-2-7B model, where the next-token prediction with the instruct model yields a notably higher score (62.6\%) compared to the full instruct benchmarking score (50.3\%).

However, the AstroLLaMA models exhibit a different pattern when comparing their full instruct scores to their base model token prediction scores. For AstroLLaMA-2-7B-AIC, the full instruct score (41.4\%) is lower than its base model token prediction score (44.3\%). AstroLLaMA-3-8B-AIC shows an even more pronounced deficit with a full instruct score of 61.8\% compared to its base model token prediction score of 71.9\%, a ten-point decrement. For AstroLLaMA-2-70B-AIC, the difference is also pronounced, with a full instruct score of 64.7\% compared to a base model token prediction score of 76.0\%.

We have attempted to improve performance by varying training hyperparameters, but without success. Our findings suggest that the current SFT dataset, inherited from the original AstroLLaMA series in \cite{perkowski2024}, appears inadequate. This is perhaps not entirely surprising, given that the current SFT set consists of only 30,000 Q\&As, many of which pertain to open-source Q\&As with little relevance to astronomy. This observation indicates that proper SFT not only requires a much larger dataset to maintain performance but also, critically, that the content of the SFT dataset is crucial for maintaining the knowledge structure without deviating the model too much towards general answers. 

In this study, we limited ourselves to the old SFT set to ensure a fair comparison with the existing AstroLLaMA series. However, we have found that increasing the Q\&A set to approximately 50 million, extracted from astronomy literature, resolves this issue. We plan to report on this development in an upcoming paper with the proper release of the latest AstroLLaMA (de Haan et al., in prep.).

The underperformance of SFT models has led us to conclude that a more appropriate comparison of potential knowledge gain from CPT might be achieved through next-token prediction, especially with the base model without SFT. This approach reveals that the AstroLLaMA-2-70B outperforms the native LLaMA-2-70B, with scores of 76.0\% and 73.9\% respectively.

As discussed in Ting et al. 2024 \cite{ting2024}, an improvement of about 3.5 points is equivalent to approximately a 10-fold increase in value when extrapolating from the current score and price trade-off of some proprietary models. The observed gain of 2.1 points is therefore quite notable, representing an important performance boost. This improvement is comparable to the two-third of the performance gain observed between models like Claude-Haiku to Claude-Sonnet or GPT-4o-mini to GPT-4o.

Moreover, the performance of AstroLLaMA-2-70B (76.0\%) begins to rival some of the flagship models available when this study was conducted. Gemini-1.5-Pro-001 achieved a score of 77.6\%, while GLM-4-0520 and Claude-3.0-Sonnet scored 75.1\% and 76.7\% respectively. These comparisons underscore the effectiveness of our CPT approach in enhancing model performance for astronomy-specific tasks, bringing the AstroLLaMA-2-70B model close to the performance levels of some of the most advanced general-purpose language models.

Interestingly, for AstroLLaMA-2-70B, the next-token prediction with the instruct model also achieves a marked improvement over the LLaMA-2-70B, scoring 75.4\% compared to 71.4\%. Although this is lower than the next-token prediction score of 76.0\% for the base model, it suggests that for larger models, the relatively small SFT set does not impede the underlying knowledge structure. Instead, it primarily affects the model's ability to follow instructions. 

However, the same pattern does not hold for the previously existing AstroLLaMA-2-7B models. Both the version trained only on abstracts \cite{nguyen2023} and the one trained on abstracts, introductions, and conclusions of astro-ph papers \cite{perkowski2024} perform worse than the native LLaMA-2-7B models. The base model token prediction scores for AstroLLaMA-2-7B-Abstract and AstroLLaMA-2-7B-AIC are 43.5\% and 44.3\% respectively, compared to 51.3\% for LLaMA-2-7B. This represents a decrement of approximately 7-8 points. These results highlight a key finding of our study: specialized LLMs might not necessarily improve performance, especially for smaller models.

For comparison with previous work, we also studied the LLaMA-3-8B series, performing the CPT and SFT processes similarly to \cite{perkowski2024}. For the model trained on abstracts, introductions, and conclusions, the next-token prediction with the base model model scores 71.9\%, which is comparable to the native LLaMA-3-8B's score of 72.0\%. However, the performance of AstroLLaMA-3-8B after SFT shows a visible decline, with next-token prediction scoring 68.4\% and full instruct scoring 61.8\%.

These results indicate that, unlike AstroLLaMA-2-7B, AstroLLaMA-3-8B appears to retain knowledge more effectively, at least before its performance is affected by the limited SFT set. Yet, it fails to improve upon the baseline score, in contrast to the 70B model. Interestingly, our preliminary findings (to be detailed in de Haan et al., in prep.) suggest that training even with the entire full text of astro-ph papers, using a cleaner version obtained through OCR, does not improve upon the baseline for the 8B model. We have found that incorporating other, more curated astronomy literature and data beyond astro-ph can improve the score even at the 8B scale. However, we will defer a full discussion of these findings to our upcoming study.

These observations are further supported by our findings with the AstroLLaMA-3-8B version trained on summaries of all full-text astro-ph papers. While this model still does not surpass the native model, its overall performance is more consistent. The base token prediction score for AstroLLaMA-3-8B-Summary is 72.3\%, slightly higher than the native LLaMA-3-8B (72.0\%). After SFT, the next token prediction score (70.9\%) and full instruct score (69.0\%) show less degradation compared to the AIC version (68.4\% and 61.8\% respectively).

These results suggest that high-quality, information-dense tokens used in CPT, combined with a substantially larger set of astronomy-focused Q\&As for SFT, might be critical for advancing lightweight models to outperform their native counterparts.

\section{Discussion and Conclusion}

The development of specialized language models for astronomical research has gained significant attention recently, driven by the vast and general reasoning abilities of LLMs. These capabilities make it possible to tackle complex tasks as research agents in astronomy \cite{sun2024b}. A key aspect of developing an automated AI Astronomer is having a reliable, robust, and lightweight LLM that can handle specialized astronomical knowledge for individual downstream tasks at scale. This challenge has sparked interest in developing specialized LLMs for astronomy.

Although some specialized LLMs have been previously published, their performance evaluation has been largely qualitative. The high cost of obtaining expert-labeled golden benchmarking datasets has hindered quantitative assessments. Our study aims to bridge this gap by utilizing high-quality, MCQ astronomy benchmarking datasets. This approach allows us, for the first time, to put specialized astronomical LLMs to a rigorous test. In addition to benchmarking the existing AstroLLaMA-2-7B, we have also trained AstroLLaMA-3-8B as well as AstroLLaMA-2-70B. To our knowledge, the latter is the first 70B parameter-level specialized LLM in astronomy. Our key findings are as follows:

\begin{itemize}
\item The existing AstroLLaMA-2-7B fails to improve upon the native baseline, instead incurring a 7-8 percentage point decrement in performance.

\item AstroLLaMA-3-8B, developed in this study, does not suffer as severe a decrement and is able to retain knowledge comparable to the baseline model. However, when trained on astro-ph data, it fails to demonstrate improvement. This suggests that CPT on astro-ph data alone offers only marginal gains for the already highly performing LLaMA-3-8B models.

\item We found that the current SFT dataset, which is relatively small and not astronomy-focused, is insufficient. This lack of astronomy-tailored SFT data further decreases the performance of specialized models when comparing the instruct model versus the base model.

\item Despite the lack of improvement in smaller models, at the 70B model level, we show that the base model of AstroLLaMA-2-70B released in this study achieved visible improvement compared to the native LLaMA-2-70B. It performs close to some flagship models available when this study was conducted. This demonstrates that training specialized LLMs can indeed be effective, though with our current astro-ph bound CPT training dataset, the benefits are limited to 70B class models.
\end{itemize}

We acknowledge several limitations in this study. Notably, our MCQ questions primarily test the ability of knowledge recall. While this is currently the only astronomy-based metric (to our best knowledge) to measure LLMs' capabilities, deploying agents requires more subtle reasoning skills beyond mere knowledge recall. Furthermore, this study commenced before the release of the LLaMA-3 series. Although we included the LLaMA-3-8B model in our training for completeness, our limited academic setting made CPT of the 70B models for the LLaMA-3.1 series infeasible within the timeline of this paper. 

These limitations highlight the complexities involved in developing effective specialized LLMs for astronomy. The 70B model, even with CPT on a relatively small dataset in this study, required $\mathcal{O}(10^3)$ GPU hours. Expanding that to the full text in astro-ph and beyond would easily necessitate $\mathcal{O}(10^4)$ to $\mathcal{O}(10^5)$  GPU hours, which is often infeasible given the limited GPU resources in many academic and physical science settings. Moreover, even if successful, deploying 70B models as agents remains quite impractical. Conversely, at the 8B class level, with the already highly performant open-source models such as LLaMA-3, the need for arduous CPT may be diminished given the limited gains demonstrated in this study.

Nevertheless, our exploration provides valuable insights and directions for improving 8B models. On CPT, improving 8B models likely requires very high-quality astronomy data beyond astro-ph. A combination of textbooks, Wikipedia, and summaries might be a potential path forward. On SFT, the current open-source SFT training set is highly insufficient. Orders of magnitude larger and more astronomy-focused Q\&A datasets might prove to be key in advancing these models.

\section{Broader Impact}

Our study presents the first comprehensive benchmark effort to evaluate specialized LLMs for astronomy based on the LLaMA series, including the introduction of AstroLLaMA-2-70B, the first 70B-parameter specialized LLM for astronomy. Specialized models like AstroLLaMA could help automate redundant research efforts, allowing human experts to focus on tasks requiring higher-level reasoning or creativity. In recent years, astronomical research has seen an increasingly larger portion of funding dedicated to instrumentation, while investment in research personnel has been limited. The fraction of successful NSF grants has drastically reduced compared to a few decades ago. This shortage of personnel, combined with the increasing amount of data being collected, poses a major problem for future research. The development of specialized LLMs represents a critical step towards addressing this challenge by enabling more comprehensive analysis of observed sources, potentially uncovering unknown phenomena. By enhancing the productivity of existing research teams, specialized LLMs could help bridge the gap between the vast amount of data collected and the limited human resources available to analyze it.

\section*{Acknowledgements}
This research was made possible through the use of computing resources provided by the Oak Ridge Leadership Computing Facility Frontier Nodes. We extend our sincere gratitude to Microsoft's Accelerating Foundation Models Research (AFMR) program for their invaluable support, which was instrumental to the realization of this study.

\bibliographystyle{IEEEtran}
\bibliography{Manuscript}

\appendix

\subsection{Example of Benchmark Questions}
\label{AppendixA}

The performance scores reported in this study are derived from a proprietary benchmarking dataset curated by our research group members. The questions were extracted from the high-quality Annual Review of Astronomy and Astrophysics using Gemini-1.5-Pro-001, one of the few LLMs with a sufficiently long context window to process typical review articles, few-shot examples, and instructions (including good and bad examples from domain experts during the initial generation).

We verified that despite the questions being generated by Gemini, it doesn't appear to have an advantage when answering them, demonstrating the robustness of the questions beyond potential information leakage. Although the questions are not perfect, with some occasionally being too vague, domain experts have reviewed hundreds of such questions and found the vast majority to be robust and challenging.

For more detailed examples and information, we refer interested readers to \cite{ting2024}. However, we provide some example questions here to illustrate their nature. The questions typically match the difficulty level of graduate school general exam questions, testing detailed expert knowledge critical for research. In the U.S. educational system, students who pass such general exams in graduate school are deemed Ph.D. candidates.

Although not used in the benchmarking, we also extracted relevant text and explanations for each answer choice when generating the questions. This additional information facilitates human inspection and verification. We are in the process of releasing this benchmarking dataset but will withhold the answer key to prevent question leakage and maintain an objective benchmark for assessing LLM capabilities in detailed astronomy Q\&A in the future.

Below are some examples of the questions:

\begin{questionbox}[boxcolor1]
\textbf{Paper ID:} 1971ARA\&A...9..127S \\

\textbf{Question:} What is the most likely range of initial masses for stars that become planetary nebulae? \\

(A) 0.5 to 1.0 solar masses \\
(B) 1.0 to 1.5 solar masses \\
(C) 1.5 to 2.0 solar masses \\
(D) 2.0 to 2.5 solar masses \\

\textbf{Correct Answer:} B \\

\textbf{Explanation:} The article discusses the spatial and velocity distribution of planetary nebulae, suggesting they originate from stars with masses similar to those in the disk population. This population typically has stars with masses between 1.0 and 1.5 solar masses. Additionally, the article mentions that the rate of planetary nebula formation aligns with the rate of white dwarf creation, further supporting this mass range as white dwarfs usually originate from stars within this mass range.
\end{questionbox}

\begin{questionbox}[boxcolor3]
\textbf{Paper ID:} 2003ARA\&A..41..645R \\

\textbf{Question:} What is the primary advantage of using weak gravitational lensing to map dark matter compared to methods that rely on the distribution of light? \\

(A) Weak lensing is less affected by dust obscuration, providing a clearer view of dark matter. \\
(B) Weak lensing directly measures the mass distribution, including dark matter, rather than just the distribution of luminous matter. \\
(C) Weak lensing is sensitive to the motion of dark matter, revealing its dynamic behavior. \\
(D) Weak lensing can probe dark matter at higher redshifts than traditional methods. \\

\textbf{Correct Answer:} B \\

\textbf{Explanation:} Unlike methods that rely on the distribution of light, which only traces luminous matter, weak lensing directly measures the total mass distribution, including dark matter. This is because the bending of light by gravitational lensing is caused by the total mass, regardless of its luminosity. This direct measurement allows for a more accurate and reliable mapping of dark matter compared to methods that require assumptions about the relationship between light and mass.
\end{questionbox}

\begin{questionbox}[boxcolor2]
\textbf{Paper ID:} 2019ARA\&A..57..617M \\

\textbf{Question:} What are the primary advantages of directly imaged spectra of exoplanets compared to transmission spectra? \\

(A) Directly imaged spectra typically have higher resolution and signal-to-noise ratio, allowing for more precise measurements of atmospheric features. \\
(B) Directly imaged spectra provide information about the planet's atmospheric composition at all orbital phases, while transmission spectra only probe the day-night terminator region. \\
(C) Direct imaging allows for the detection of exoplanets at much smaller orbital separations than transit spectroscopy. \\
(D) Directly imaged spectra are less affected by the presence of clouds and hazes in the planet's atmosphere, leading to more accurate abundance estimates. \\

\textbf{Correct Answer:} A \\

\textbf{Explanation:} The article explains in section 2.3 that directly imaged spectra often have higher resolution and signal-to-noise ratio due to the use of large-aperture ground-based telescopes with adaptive optics. This allows for more precise measurements of atmospheric features. However, directly imaged spectra are typically obtained at a single, unknown orbital phase and are more affected by degeneracies due to the unknown mass, radius, and gravity of the planet.
\end{questionbox}

\subsection{Full Instruct Benchmarking Method}
\label{appendixB}

The ``full instruct" benchmarking method follows the approach proposed in \cite{ting2024}. This method is the most natural way to test an LLM's ability to function as an assistant model, as it requires question-answering in a natural language format with proper output JSON.

Below, we present the exact prompt used. The last paragraph of the prompt deviates slightly from \cite{ting2024}, as we found that repeating the exact instructions helps improve performance in terms of instruction following, especially for the current AstroLLaMA series where the proper SFT set is still somewhat lacking.

When the system allowed for a separate system prompt, we used the first paragraph (starting with ``You are an expert in general astrophysics...") as the system prompt.

The key elements that consistently improved performance across models were:
\begin{enumerate}
\item Role-play framing the task as coming from an expert in astrophysics.
\item Requesting both an answer and an explanation, encouraging a chain-of-thought inference.
\item Providing a clear output format.
\end{enumerate}

Although the third point was not always adhered to, which required us to further parse the results with GPT-4o.

The exact prompt is shown below:

\begin{questionbox}[grey]
\textbf{Prompt:}

You are an expert in general astrophysics. Your task is to answer and explain the following multiple-choice question on astrophysics, sourced from a dataset. The question is: \\

\textbf{Question}: [Question text] \\

Options: \\
\textbf{A}: [Option A] \\
\textbf{B}: [Option B] \\
\textbf{C}: [Option C] \\
\textbf{D}: [Option D] \\

Determine the correct answer using your astrophysics knowledge and provide a detailed explanation for why this answer is correct. \\

Ensure your explanation is thorough, clearly articulating your thought process based on astrophysical principles. \\

\textbf{Output format}: \\

\{ \\
    ``ANSWER": ``[The choice you decide to choose]", \\
    ``EXPLANATION": ``[Provide a valid explanation for the answer mentioned in ANSWER]" \\
\} \\

Give only one answer, either A, B, C or D, but not more than one, and always give an answer.

Provide your response in valid JSON format only. Begin your output with the JSON structure immediately, without any preceding text. Strictly adhere to the specified output format.
\end{questionbox}

As we have seen in the main text, while this full instruct method performs well for the native LLaMA series models, the lack of a proper SFT set seems to cause the AstroLLaMA models to underperform, despite their improved knowledge base (e.g., in AstroLLaMA-2-70B). This observation prompted us to also consider next-token prediction as described in the following sections.

\subsection{Next Token Benchmarking Method}
\label{appendixC}

To better assess if the CPT can lead to improved base model performance without potential bias from the SFT set, we also test the base models. However, base models, by definition, are not capable of following exact instructions as they are only trained to perform next-token prediction.

Therefore, testing the base model (post-CPT) requires structuring the question in a form where the next token after the ``question" has a high probability of being one of the potential answers: A, B, C, or D from the MCQ. The MCQ benchmarking is advantageous in this way as it only requires understanding the output of the next token, and here we chose the answer with the highest probablity (logit) as the suggested answer of the LLM.

We structure the benchmarking prompt as follows:

\begin{questionbox}[grey]
\textbf{Prompt:}\\
Astrophysics and Cosmology\\
Multiple choice questions \\
Solution set:\\
\\
\textbf{Question}: [Example Question 1]\\
A: [Option A]\\
B: [Option B]\\
C: [Option C]\\
D: [Option D]\\
\\
\textbf{Answer}: [Correct Answer]\\
\\
\textbf{Question}: [Example Question 2]\\
A: [Option A]\\
B: [Option B]\\
C: [Option C]\\
D: [Option D]\\
\\
\textbf{Answer:} [Correct Answer]\\
\\
\textbf{Question:} [Actual Test Question]\\
A: [Option A]\\
B: [Option B]\\
C: [Option C]\\
D: [Option D]\\
\\
\textbf{Answer:}
\end{questionbox}

Key aspects of this prompting strategy include:

\begin{enumerate}
\item We provide two example questions with their correct answers before presenting the actual test question. This gives the model a clear pattern to follow, which is particularly beneficial for base models not specifically fine-tuned for instruction following.

\item The format for each question, including the example questions and the test question, is kept consistent. This helps the model recognize the pattern and predict the next token more accurately.

\item The prompt ends with ``Answer:", encouraging the model to predict the next token as one of the multiple-choice options (A, B, C, or D).

\item We account for variations in token representation across different models. For instance, some models may represent answer choices as ``A", ``B", ``C", ``D", while others use `` A", `` B", `` C", `` D" (note the leading space). Our evaluation dynamically identifies the correct token representation for each model.
\end{enumerate}

This method allows us to evaluate the base model's knowledge and reasoning capabilities in a way that aligns with its pre-training objective, providing insights into the effectiveness of the CPT process without the potential confounding effects of SFT.

Nonetheless, since the next-token prediction can also be applied to the instruct model (post SFT), we apply this method to three scenarios for completeness: the full instruct benchmarking method (as described in Appendix B), next-token prediction for the base model, and next-token prediction for the instruct model.

This comprehensive approach allows us to disentangle the effects of CPT and SFT on model performance, providing a more nuanced understanding of how each stage of training impacts the model's ability to answer astronomy-related MCQs.

\end{document}